\documentclass[aoas,preprint]{imsart}

\RequirePackage[OT1]{fontenc}
\RequirePackage{amsthm,amsmath}
\RequirePackage{natbib}
\RequirePackage[colorlinks,citecolor=blue,urlcolor=blue]{hyperref}
\usepackage{graphics,graphicx}
\usepackage{longtable}
\usepackage{epsfig}
\usepackage{wrapfig}
\usepackage{color,lscape}
\usepackage{multirow,multicol}
\usepackage[pagestyles]{titlesec}
\usepackage[normalem]{ulem}
\usepackage{bm}
\arxiv{arXiv:1810.05297}

\startlocaldefs
\numberwithin{equation}{section}
\theoremstyle{plain}

\endlocaldefs
\begin{document}

\begin{frontmatter}
\title{Bayesian Hierarchical Spatial Model for Small-area Estimation 
with Non-ignorable Nonresponses and Its Application to the NHANES Dental Caries Data}
\runtitle{Bayesian Model for Dental Survey Data}
%\thankstext{T1}{Footnote to the title with the ``thankstext'' command.}

\begin{aug}
\author{\fnms{Ick Hoon} \snm{Jin}\thanksref{t1,m1}\ead[label=e1]{ijin@yonsei.ac.kr}},
\author{\fnms{Fang} \snm{Liu}\thanksref{t1,m2}\ead[label=e2]{Fang.Liu.131@nd.edu}},
\author{\fnms{Evercita} \snm{Eugenio}\thanksref{m2}\ead[label=e3]{eeugenio@nd.edu}},
\author{\fnms{Kisung} \snm{You}\thanksref{m2}\ead[label=e4]{kyou@nd.edu}},
\and
\author{\fnms{Suyu} \snm{Liu}\thanksref{m3}\ead[label=e5]{syliu@mdanderson.org}
}

\thankstext{t1}{Jin and Liu are co-first author.}
\runauthor{Jin et al.}

\affiliation{Yonsei University\thanksmark{m1}, University of Notre Dame\thanksmark{m2}, and \\
University of Texas MD Anderson Cancer Center\thanksmark{m3}}

\address{Department of Applied Statistics\\
Yonsei University\\
Seoul. Republic of Korea. 03722\\
\printead{e1}}

\address{Department of Applied and Computational\\
Mathematics and Statistics\\
University of Notre Dame\\
Notre Dame. IN. USA. 46556\\
\printead{e2}\\
\phantom{E-mail:\ }\printead*{e3}\\
\phantom{E-mail:\ }\printead*{e4}}

\address{Department of Biostatistics\\
The University of Texas\\
MD Anderson Cancer Center\\
Houston. TX. USA. 77030.\\
\printead{e5}}
\end{aug}

\begin{abstract}
The National Health and Nutrition Examination Survey (NHANES) is a major program of the National Center for Health Statistics, designed to assess the health and nutritional status of adults and children in the United States. The analysis of NHANES dental caries data faces several challenges, including (1) the data were collected using a complex, multistage, stratified, unequal-probability sampling design; (2) the sample size of some primary sampling units (PSU), e.g., counties, is very small; (3) the measures of dental caries have complicated structure and correlation, and (4) there is a substantial percentage of nonresponses, which are expected not to be missing at random or non-ignorable. We propose a Bayesian hierarchical spatial model to address these analysis challenges. We develop a two-level Potts model that closely resembles the caries evolution process, and captures complicated spatial correlations between teeth and surfaces of the teeth. By adding Bayesian hierarchies to the Potts model, we account for the multistage survey sampling design, while also enabling information borrowing across PSUs for small-area estimation. We incorporate sampling weights by including them as a covariate in the model and adopt flexible B-splines to achieve robust inference. We account for non-ignorable missing outcomes and covariates using the selection model. We use data augmentation coupled with the noisy Monte Carlo algorithm to overcome the numerical difficulty caused by doubly-intractable normalizing constants and sample posteriors. Our analysis results show strong spatial associations between teeth and tooth surfaces, including that dental hygienic factors, such as fluorosis and sealant, reduce dental disease risks. 
\end{abstract}

% \begin{keyword}[class=MSC]
% \kwd[Primary ]{60K35}
% \kwd{60K35}
% \kwd[; secondary ]{60K35}
% \end{keyword}


\end{frontmatter}

\section{Introduction} \label{sec:Introduction}
The National Health and Nutrition Examination Survey (NHANES) is a major program of the National Center for Health Statistics  and focuses on understanding the health and nutrition of adults and children in the United States. 
%NHANES is composed of two parts. The first is a survey that is administered each year to a random sample of the population (about 5,000 people); participants are asked a set of demographic, socioeconomic, dietary, and health-related questions. The second part consists of a physical exam that includes medical, dental, physiological measurements, and laboratory testing administered to people located in 15 selected counties across the country. 
The data collected from this survey is used to determine the health status of Americans, developing nutritional guidelines, and forming better health policies \citep{nhanesgeneral}. 
In this paper, we analyze the NHANES dental caries data to understand the relationship between the demographic or dental hygienic factors and dental caries.

%Dental caries, also known as tooth decay, is one of the most prevalent chronic diseases worldwide \citep{selwitz2007dental}. Although preventable, people remain susceptible to the disease throughout their lifetime (\citealp{Featherstone:2000p887}, \citealp{pitts2004we}); hence it remains a major global oral health burden and is prevalent in the United States. Caries are triggered by acids produced during bacterial fermentation of food debris that accumulates on the tooth surface. This causes localized dissolution of the tooth's hard tissues and leads to the development of cavities or holes in the teeth \citep{Kidd:2003}. The four main factors influencing the formation of dental caries are the person's age, the health of the tooth surface, the presence of cariogenic bacteria, and the presence of fermentable carbohydrates \citep{Soames:1993}. The degree of caries progression varies by individual, depending on the shape of the teeth, oral hygiene habits, and the buffering capacity of saliva. If left untreated, caries can spread to supporting tissues and the jaws, and result in advanced conditions that are often painful \citep{Jamison:2006} and may lead to tooth loss. In this analysis, we focus on two major dental caries outcomes -- presence vs. absence of teeth at the tooth level, and healthy vs. non-healthy tooth surfaces at the surface level, and aim to identify the demographic or dental hygienic factors that relate to the dental caries outcomes. 

There are several challenges to the analysis of the NHANES dental caries data: (1) the data were collected using a complex, multistage, stratified, unequal-probability sampling design. It is important to incorporate the sampling design feature, as well as sampling weights, into the model and inference \citep{Breidt:2000, Zheng:2003p99, Zheng:2004p209, Zheng:2005p1, Opsomer:2005, Chen:2010p23, Zhang:2015}. (2) Some of the primary sampling units (PSU), e.g., counties, have very small sample sizes, making PSU-level inference highly unreliable or sometimes impossible. This is known as the ``small-area estimation'' problem in the survey sampling literature \citep{rao2015small}. (3) The collected data have a complicated structure and correlation. The outcomes consist of tooth-level measurements and also (tooth) surface-level measurements, where the surface measurements are nested within the tooth-level measurements and they are spatially correlated \citep{Garcia:2007p3223}. For example, the health status of a surface on a particular tooth might be influenced by the disease status of proximal surfaces or teeth, and the absence of a tooth might relate to the absence/presence of nearby teeth. (4) There are a substantial number of nonresponses in both outcomes and covariates (e.g., household income), and the resulting missing data are potentially non-ignorable.  

To address these challenges, we develop a Bayesian hierarchical spatial model for small-area estimation with non-ignorable nonresponse. We account for the multistage sampling scheme using the Bayesian hierarchical model structure, which also enables information borrowing across PSUs to improve the efficiency of small-area estimation. We incorporate sampling weights by including them as a covariate in the model and adopt flexible B-splines to achieve robust inference. We capture the feature that the surface measurements are nested within the tooth-level measurements, and they are spatially correlated by using a two-level spatial model. At the first level of hierarchy, the trinary probability of a tooth being present, absent due to the dental disease, or absent due to other reasons is modeled via a Potts model. Conditional on the tooth being present, we model the probability of a decayed, filled or healthy surface via a second Potts model. We employ the selection model to account for non-ignorable missing outcomes and covariates. Estimation of the proposed Bayesian hierarchical spatial model is challenging, because of the presence of a doubly-intractable normalizing constant and non-ignorable missing data. We use the noisy Monte Carlo algorithm, coupled with data argumentation, to make posterior inference.

There is a rich body of literature on modeling caries outcomes. \citet{Garcia:2007p3223} analyzed the caries experience data with the conditionally specified logistic regression model \citep{Joe:1996p113} and a multivariate probit model \citep{Chib:1998p347}. \citet{Afroughi:2010p374} modeled caries of deciduous teeth in children using the spatial autologistic regression model. % and identified a risk pattern of decayed dents in these children's teeth. 
\citet{Bandyopadhyay:2011p85} developed a multivariate spatial beta-binomial model for the total count of decayed, missing, or filled surfaces in a tooth.
% which accommodates both over-dispersion and latent spatial associations. 
\citet{Mustvari:2013p5241} used a spatially referenced multilevel autologistic model to  analyze caries data.
%investigate (i) if caries experience outcomes recorded at surface level were spatially associated; and (ii) if the dental examiners exhibited some spatial behavior while scoring caries experience at surface level. 
\citet{Jin:2016p884} developed a Bayesian hierarchical two-level framework that closely resembles the caries evolution process in humans.

Limited research has been done for dental data collected from a survey for small-area estimation \citep{ghosh1994small, rao2015small}.
%For dental data that was collected from a survey with small sub-populations, various small-area estimation techniques have been employed for parameter estimation \citep{ghosh1994small, rao2015small}. %\citet{leroux1996estimation} is one of the first papers to use small-area estimation on dental data. Specifically, the data was collected from an oral health survey in the state of Washington, and small area estimation was used for analyzing dental disease and sealant use. However, this early study was small in scale, and did not include any spatial components. 
\citet{antunes2002spatial} used small-area estimation and spatial models to describe the epidemiological measurements collected from small sub-population (districts in Sao Paulo, Brazil), and to examine the association between tooth decay and dental treatments in children. However, the spatial analysis was applied to the geographic districts and did not consider the spatial correlations among teeth or surfaces. \citet{gentili2015small} developed a small area estimation spatial model to analyze access to pediatric primary care. Similarly, the spatial model was with regard to the geographic location of the pediatric primary care, rather than on dental data per sample. 

The remainder of the paper is structured as follows. In Section 2, we describe the NHANES study design and the dental caries data. In Section 3, we present the exploratory data analysis results that help us to build models for the data. In Section 4, we propose a Bayesian hierarchical spatial model for the outcomes, which incorporates the sampling design and weights, while accounting for the non-ignorable missing covariates and outcomes. 
%We present the Bayesian computational framework using the noisy Monte Carlo algorithm \citep{Liang:2013p2199, Alquier:2016p29} in Section 5. 
In Section 5, we apply the Bayesian model to the NHANES dataset and summarize analysis results. We provide our conclusions and future developments in Section 6.

\section{NHANES Dental Caries Data}\label{sec:data_describe}

The NHANES dental caries data analyzed were collected from 1999 to 2004. This is the latest publicly available data focusing on young adults, aged 20 to 34 years, and providing information on PSU, sampling weights, and the dental covariates of interest (e.g., sealant and fluorosis). More recent NHANES data, such as 2015-2016 data, did not collect information on fluorosis, an important factor affecting dental health. The 1999-2004 NHANES dental caries data were collected over 87 PSUs via a complex, multistage and stratified sampling design. The sampling procedure consisted of four stages: 
\begin{itemize}
\item Stage 1: PSUs, mostly single counties, are selected with probability proportional to its size (PPS), from strata defined by geography and proportions of minority populations. %We denote the probability that PSU $i$ is sampled by $p_{i}$. 
\item Stage 2: The PSUs are divided into segments, generally city blocks or their equivalents. The segments are sampled with PPS. %We denote the probability of segment $j$ in PSU $i$ being sampled by $p_{j|i}$. 
\item Stage 3: A household is randomly sampled from each segment selected in Stage 2. NHANES over-sampled certain subgroups of particular public health interest. In the 1999-2004 surveys, African Americans, Mexican Americans, and persons age 60+ are examples of over-sampled subgroups. %We denote the probability that household $k$ in segment $j|i$ is sampled by $p_{k|ji}$. 
\item Stage 4: Individuals in a selected household in Stage 3 are randomly chosen from designated age-sex-race/ethnicity screening sub-domains. In other words, all eligible members in a household were listed, and a sub-sample of individuals was chosen based on sex, age, and race or ethnicity. %We denote the probability that individual $l$ in household $k|ji$ is sampled by $p_{l|kji}$.
\end{itemize}
%The base sampling weight is first calculated for each individual in the survey, incorporating the above 4 sampling stages, and is proportional to the inverse sampling probability $(p_ip_{j|i}p_{k|ji}p_{l|kji})^{-1}$. 
The NHANES data has a weight variable that is inverse proportional to each individual being sampled to the survey. The details on the calculation of the weight can be found at the NHANES official website \url{https://wwwn.cdc.gov/nchs/nhanes/tutorials/module3.aspx/}. In a nutshell, a base weight is first calculated based on the above 4 sampling stages, which is then adjusted for non-response in the in-home interview and the physical exam (including the dietary interviews, body measurements, blood work except for young children, dental exam, and other tests). Finally, it is post-stratified to match the 2000 US Census population total for each sampling sub-domain to obtain the final weight. 

Survey participants either received an oral dental exam or not. Those who received an oral dental exam and did not have any teeth or any molar teeth were excluded from the analysis. The number of respondents who participated in both the oral health survey and the oral exam is 3595; and the number of participants who only answered the oral health survey is 321, leading to a total number of 3916 participants for our analysis.

% \begin{figure}
% \scalebox{0.9}{
% \colorbox[rgb]{1,.95,.89}{
% 	\begin{tabular}{c}
% 	\includegraphics[scale=0.5]{TopRowCrop.eps}\\
% 	\centerline{\includegraphics[scale=0.5]{Lingual.eps}\hspace*{.5in}
% 	\includegraphics[scale=0.5]{Occlusal.eps}}
% 	\end{tabular}
% }
% }
% \caption{\label{fig:Dental}
% Different surfaces of permanent dentition within a human mouth (adapted from \citealp{Darby:1995})}
% \end{figure}

% Figure \ref{fig:Dental} illustrates the different surfaces of permanent dentition within a human mouth. 

Following \citet{Darby:1995}, the entire dentition can be divided into four quadrants: two on each jaw bone, the mandible (lower jaw) and maxilla (upper jaw). Each quadrant consists of a cluster of seven teeth, excluding wisdom teeth: the non-anterior teeth (two molars and two premolars) and the anterior teeth (one incisors and two canine). In the study of dental caries, each non-anterior tooth contributes five surfaces (occlusal, mesial, distal, facial, and lingual), while each anterior tooth contributes four of these surfaces, lacking an occlusal surface. Because dental data consist of a two-level hierarchy (i.e., a tooth level and a surface level), the primary response variable is recorded differently according to the level of hierarchy. An assessment of the current status for caries progression at the tooth level is a trinary indicator for the presence of a particular tooth, absence of a particular tooth due to dental disease, and absence of a particular tooth due to other reasons. Next, conditional on the tooth being present, an assessment of the current status of caries progression at the surface level of a tooth is a trinary indicator that each surface is either healthy (H), decayed (D), or filled (F). If the whole tooth is missing, then all the surfaces are considered missing. The reason for a missing tooth was determined from the questionnaire administered to the study participants. We acknowledge that this self-reported information may be inaccurate, but it is the best information available. 

Several individual-level covariates also were collected, including gender (0 = male, 1 = female), poverty-income ratio (0 = below poverty line, 1 = above poverty line), race (1 = non-Hispanic white; 2 = non-Hispanic black; 3 = other races, including Mexican American and Hispanics). Across all PSUs, females participated at a slightly higher rate (57\%) than males, and approximately 22\% of survey participants were below the poverty line. About 8\% of participants declined to answer the questions regarding income/poverty. 43.4\% of participants were non-Hispanic white, 20.0\% of participants were non-Hispanic black, and the rest were other races (including 26.4\% Mexican American), which was coded as the reference group in our analysis. There are two tooth-level covariates: sealant for occlusal teeth (1 = if there is a sealant, 1 = if there is no sealant) and fluorosis level for each tooth (0 = normal, 1 = very mild, 2 = mild, 3 = moderate, and 4 = severe). If a participant did not receive an oral dental exam, then the sealant and fluorosis information is also missing, along with the outcome data. 

In the subsequent analysis, we standardized the covariates using the method proposed by \citet{Gelman:2008p1360}. Specifically, binary inputs were shifted to have a mean of 0 and to differ by 1 in their lower and upper conditions. For example, since female respondents in our study are 57\% and male are 43\%, we define the centered ``gender'' variable to take on the values 0.43 and -0.57. Other inputs were shifted to have a mean of 0 and scaled to have a standard deviation of 0.5. 

\section{Exploratory Data Analysis} \label{sec:EmpiricalStudies}
We performed exploratory data analysis on the data to guide us through the development of models for the outcome variables (presence vs. absence of teeth and healthy vs. unhealthy tooth surface for each PSU), the incorporation of the sampling weights in the models, and the development of the selection models in order to model the missingness in the outcome and covariates with missing values. 

\subsection{Sampling Weights} \label{sec:Sampling_Weights}

To incorporate sampling weights in the data analysis, we employed the model-based approach by including the weight as a covariate in the model of the outcome variables. Model-based survey data analysis with sampling probabilities as covariates is  well established  \citep{Zheng:2003p99, Zheng:2004p209, Zheng:2005p1, Chen:2010p23} and leads to consistent and efficient estimates for the predicted values of survey variables under the assumed model \citep{little2004}.  The weight in the NHANES data has a wide range $[465.59, 69220.78]$ across the PSUs that is on a different magnitude scale from the rest of covariates. We therefore standardized the weight so that it has a mean of 0 and a standard deviation of 0.5. 

The key issue is how to determine the function form between the caries outcomes and sampling weights. We plotted the logit of the proportion of surfaces with cavities in an individual vs. the individual's weight by PSU, and then applied the locally weighted scatter-plot smoothing (LOWESS) to smooth out the relationship. The plug-in bandwidth selection method \citep{Sheather:1991p683} was used to select the bandwidth for the LOWESS. 

As shown in Figure 1 in Section B of the Supplement Materials,  the relationship between the caries outcomes and sampling weights is nonlinear and varies across PSU.
%differs by PSU, and also has various nonlinear trends.
%Figure \ref{fig:lowess} shows the relationship from a few PSUs (the rest of the plots are available in Section B of the supplementary materials). This analysis suggests that the relationship differs by PSU, and also has various nonlinear trends. 
We observed similar nonlinear pattern when plotting the logit of the proportion of absent teeth in an individual vs. weight (see Figure 2 in Section B of the Supplement Materials). Therefore, we employ the nonparametric regression with B-splines to model the curvature relationship between the caries outcomes and weight. \citet{Breidt:2000} previously used a nonparametric regression approach (e.g., smoothing spline) to incorporate sampling weights for analyzing survey data. 

%\begin{figure}
%\centering
%\includegraphics[scale=0.3]{Weight10.pdf}
%\includegraphics[scale=0.3]{Weight12.pdf}
%\includegraphics[scale=0.3]{Weight13.pdf}
%\caption{\label{fig:lowess}
%LOWESS plots between the logit of the proportion of surfaces with cavity vs. weight }
%\end{figure}

\subsection{Small Area Estimation Issue} \label{sec:smallarea}
%Table \ref{table:samples} lists the sample sizes in the PSUs. 
As summarized in Table 1 in Section B of the Supplement Materials, most of the PSUs have $30\sim 69$ subjects, 8 PSUs have $20 \sim29$ subjects, and the PSU with the largest sample size has 84 participants. Small sample size makes statistical inference challenging in some PSUs, known as small area estimation problem \citep{rao2015small}. Following \citet{rao2015small}, we employ Bayesian hierarchical model to borrow information across the PSUs (i.e., small areas) to overcome this problem. 

To guide the choice of an appropriate Bayesian hierarchical model,  we conducted a preliminary analysis of the two outcome variables based on the model described in Section \ref{sec:MeasurementModel}, but without considering hierarchical model structure and missing data, in each of the 87 PSUs separately. % and only with complete cases. 
Due to sparse data, out of 87 PSUs, only 68 were estimable. Figure 3 in Section B of the Supplement Materials shows the density of the estimated regression coefficients obtained from these 68 PSUs. The empirical distributions of each estimated coefficients across the PSUs are roughly bell-shaped, providing some empirical evidence for the adoption of a Gaussian hierarchical model, which assumes that PSU-level parameters follow a Gaussian distribution.

\subsection{Missing values in outcome and covariate} 
\label{sec:Missing}

%We conducted a preliminary analysis of the two outcome variables based on the model described in Section \ref{sec:MeasurementModel}, but without considering hierarchical model structure and missing data, in each of the 87 PSUs separately and only with complete cases. Due to sparse data, out of 87 PSUs, only 68 were estimable. Figure \ref{fig:density} shows the density of the estimated regression coefficients obtained from these 68 PSUs. The empirical distributions of each estimated coefficients across the PSUs are roughly bell-shaped, providing some empirical evidence for the adoption of a hierarchical model, assuming PSU-level parameters follow a Gaussian distribution.

The NHANES data has two major sources of missing data: (1) subjects that elected not to provide family income information in the survey, which is used to calculate ``poverty,'' referred to as non-responders in poverty (NORP); (2) subjects that did not take the dental exam, referred to non-responders in dental exam (NORD). As shown in Figure 4 in Section B of the Supplement Materials, 
%The histograms on the missing percentages of NORP and NORD are given in Figure \ref{fig:misealflu}. The 
the NORP missing percentage ranges from 0\% to 31\% across the 87 PSUs, while the NORD ranges from 0\% to 24\%. 

%Only 18 PSUs out of 87 do not have NORP; 6 PSUs do not have NORD. 

%\begin{figure}
%\begin{center}
%%\begin{tabular}{cc}
%\includegraphics[scale=0.45]{missingpovertybypsu.pdf} 
%\includegraphics[scale=0.45]{missingsealantfluorosisbypsu.pdf} 
%%\end{tabular}
%\end{center}
%\caption{\label{fig:misealflu}(a) Percentage of NORP for each PSU and (b) percentage of NORD for each PSU.]}
%\end{figure}

It is widely known that income, if collected in a survey, is subject to missingness not at random (MNAR) -- compared with respondents who reported income, respondents with missing income information generally appeared younger, less educated, and of lower parity \citep{Kim:2007}. The NORD subjects have missing values in the areas of surface cavity and absence/presence of teeth, as well as in the covariates of sealant and fluorosis in NORD. These missing values are likely to be non-ignorable in the sense that not having a dental exam might correlate with the individual's oral health status, and thus the outcomes of interest (i.e., surface with cavity and absence tooth, or not). People who have very good or very bad oral health might not feel it as necessary to go to the oral exam.
To account for these non-ignorable missing data, it is imperative to model the missing data mechanism (Little and Rubin, 2002). In what follows, we first describe our measurement model, and then the missing-data model. 

\section{Methodology} \label{sec:Methdology}
\subsection{Measurement Model} \label{sec:MeasurementModel}
In this section, we introduce a Bayesian hierarchical spatial model that accommodates the spatial interactions in dental structures and the sampling weights from the complex design. At the tooth level, we consider only one spatial interaction (i.e., the interaction with neighboring teeth). We denote the corresponding parameter for this interaction as $\psi_t \in [0, \infty)$. At the surface level, we consider three types of spatial interactions: 1) non-occlusal surfaces on the same tooth (type-A interaction), 2) surfaces on adjacent teeth on the same jaw (type-B interaction), and 3) contact surfaces on the opposite jaw (type-C interaction). For the sake of simplicity and ease of interpretation, we eliminated the interactions between non-neighboring surfaces, such as the interaction between the facial and lingual surfaces of the same tooth. 

\begin{figure}
\begin{tabular}{cc}
(a) & (b) \\
\includegraphics[width=0.43\textwidth]{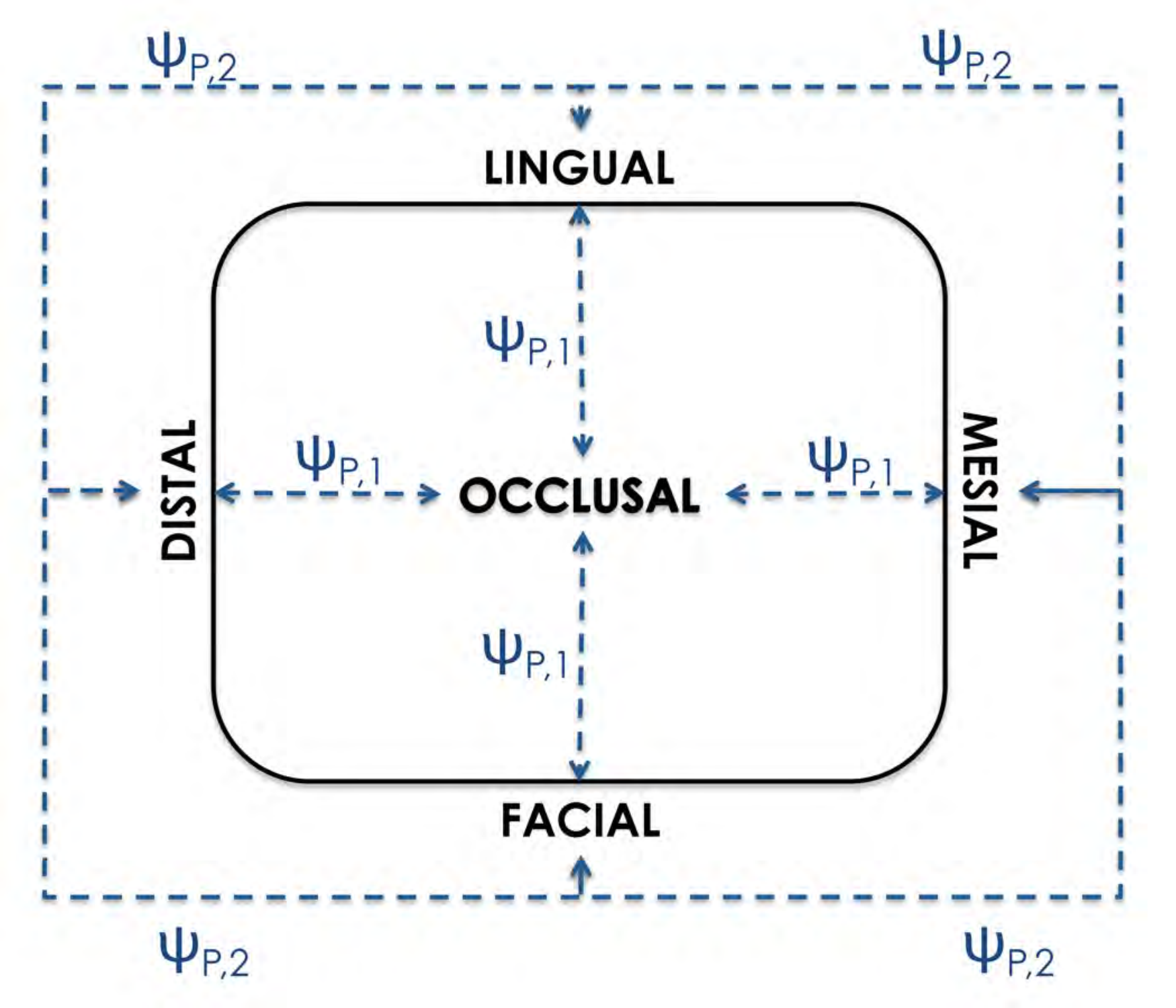} &
\includegraphics[width=0.46\textwidth]{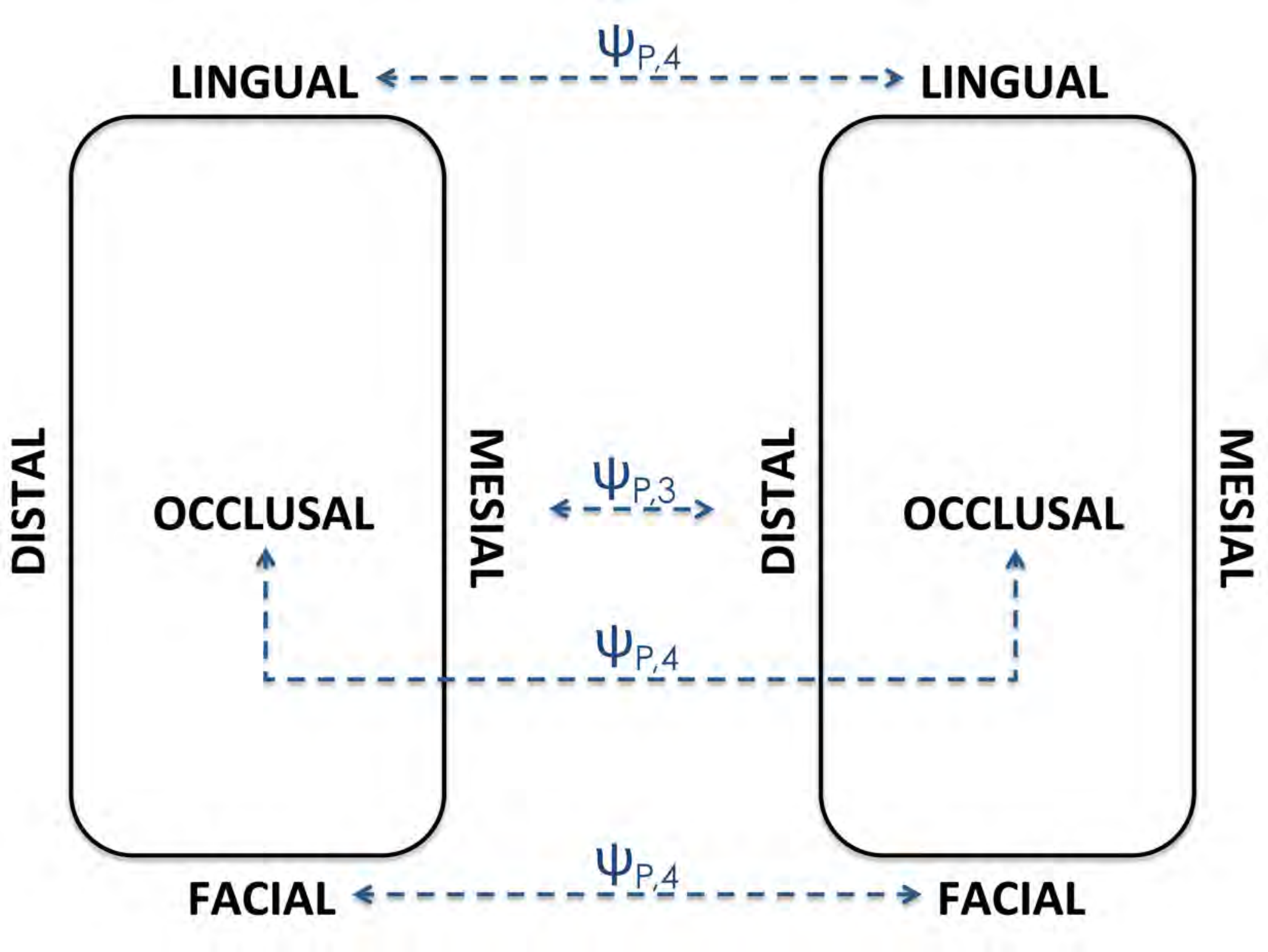} \\
\end{tabular}
\caption{
Illustrations of spatial interactions at the surface level from \citet{Jin:2016p884}. Panel (a) denotes the type-A (within-tooth) interactions; panel (b) represents the type-B (between-teeth) interactions.}\label{fig:DMFS}
\end{figure}

We illustrate all the different types of interactions in Figure \ref{fig:DMFS}. Type-A interactions consist of two categories, characterized by two spatial interaction parameters, $\psi_{p,1}$ and $\psi_{p,2}$. Specifically, $\psi_{p,1}$ denotes associations between the occlusal surface and the other four surfaces on the same tooth, while $\psi_{p,2}$ denotes associations between adjacent non-occlusal surfaces on the same tooth (i.e., between mesial - facial, mesial - lingual, distal - facial, and distal - lingual surfaces). 
Type-B interactions consist of two categories characterized by two spatial parameters, $\psi_{p,3}$ and $\psi_{p,4}$. While $\psi_{p,3}$ denotes interactions between the contacting mesial and distal surfaces of adjacent teeth on the same jaw, $\psi_{p,4}$ quantifies the interactions between adjacent occlusal surfaces, facial surfaces, and lingual surfaces of teeth on the same jaw. Finally, $\psi_{p,5}$ is the parameter that captures the spatial correlation between the contacting occlusal surfaces on opposite jaws (Type-C interaction). We denote the vector of the spatial association parameters by $\psi_p = \{\psi_{p,1}, \cdots, \psi_{p,5}\}$ where $\psi_{p,1}, \cdots, \psi_{p,5} \in [0, \infty)$. The defined spatial interactions $\psi_{t}$ and $\psi_p$ are incorporated in the Potts models for tooth and surface outcomes. We may consider a multinomial framework in the spatial generalized linear models (SGLM) 
that uses a latent Gaussian Markov random field to model spatial dependence. However, for SGLM, 
it is difficult to interpret spatial dependence and choosing the cut-off values for the latent Gaussian Markov random fields is challenging because dental outcomes are nominal values.

\subsubsection{Model for the Presence/Absence of Teeth for Each PSU}
\label{sec:Potts}
Recall that $x_{ijk}$ denotes the trinary variable indicating whether the $k$th tooth of $j$th individual at $i$th PSU is absent due to the dental disease $(x_{ijk} = 2)$ ($m$), absent not due to the dental disease $(x_{ijk} = 3)$ ($\bar{m}$), or present $(x_{ijk} = 1)$ with $i = 1, \cdots, I$, $j = 1, \cdots, n_{i}$, and $k = 1, \cdots, 28$. We assume that ${\bf x}_{i} = \{x_{ijk}\}$ follows a multinomial distribution via the following Potts model,

\begin{equation}\label{eq:potts}     %%%%%%%%%%%%%%%%%%%%%%%
\begin{split}
f\Big({\bf x}_{i} \mid \boldsymbol\theta_{t_i}\Big) &= \frac{1}{\kappa(\boldsymbol\theta_{t_i})}
\exp\bigg[ \psi_{t_i} \sum_{(jk) \sim (jk)'} I\Big(x_{ijk} = x_{(ijk)'}\Big) \\
&+\! \sum_{(j,k)} I(x_{ijk} = 2) \Big\{\alpha_{m_i} + \sum_{r=1}^6 \beta_{r,m_i} z_{r,ij} \Big\}\\
&+\!\sum_{(j,k)}\!I(x_{ijk} = 3)\! \Big\{\alpha_{\bar{m}_i}\! +\! \sum_{r=1}^6 \beta_{r,\bar{m}_i}z_{r,ij} \Big\} \\
&+\! \sum_{(j,k)} \!\Big\{I(x_{ijk} = 2) \!+\! I(x_{ijk} = 3)\Big\}\!\sum_{q=1}^{k+2} \beta_{q,t_i} B_q(\pi_{ij}) \bigg]\!\!\!\!
\end{split}
\end{equation}
where $\psi_{P_i}$ determines the intensity of interaction between $x_{ijk}$ and its neighbors, represented by $(jk) \sim (jk)'$, at the $i$th PSU; $z_{r,ij}$ is the $r$th individual-level covariate, with $r = 1, \cdots, 6$ denoting gender, poverty level, race (non-Hispanic white), race (non-Hispanic black), sealant (the binary indicator of having sealants in each individual's eligible teeth) and fluorosis (the mean of all fluorosis values for all teeth that are present); $\beta_{r,m_i}$ and $\beta_{r,\bar{m}_i}$ measure the effect of covariate $r$ for the missing teeth due to the dental disease and those not due to the dental disease, with $\alpha_{m_i}$ and $\alpha_{\bar{m}_i}$ as the intercepts, respectively; $\pi_{ij}$ is the inclusion probability (i.e., the inverse of sampling weight) for individual $j$ at the $i$th PSU; and, $B_q(\pi_{ij})$ is the quadratic B-spline basis function for the inclusion probability with its corresponding parameter $\beta_{q,t_i}$. We chose to use B-splines to model the effect of sampling weights on $x_{ijk}$, because the basis functions of B-splines are linearly independent \citep{Hastie:1992}, and thus mitigate the multicollinearity issue. In this Potts model, $\kappa\big(\boldsymbol\theta_{t_i}\big)$, with $\boldsymbol\theta_{t_i} = \big\{\psi_{t_i}, ~\alpha_{m_i}, ~\alpha_{\bar{m}_i}, ~\boldsymbol\beta_{m_i} = \{\beta_{r,m_i}\}, ~\boldsymbol\beta_{\bar{m}_i} = \{\beta_{r,\bar{m}_i}\}, ~\boldsymbol\beta_{q,t_i} = \{\beta_{q, t_i}\}\}$ is a doubly-intractable normalizing constant, which involves the sum over all possible realizations of ${\bf x}_i$, and the normalizing constant itself is a function of the parameters. Because of this doubly-intractable normalizing constant, the standard Markov Chain Monte Carlo (MCMC) algorithm cannot be applied to fit the model. We employ a new algorithm to handle this issue, which is given in the Section A of the Supplement Materials. 

\subsubsection{Model for the Health/Non-health Surface for Each PSU}\label{sec:Autologistic}

As defined previously, $y_{ijks}$ is the binary variable of the surface condition on a non-missing tooth $(x_{ijk} = 1)$, indicating whether the $s$th surface of $k$th tooth of $j$th individual at the $i$th PSU is either healthy $(y_{ijks} = 1)$, decayed $(y_{ijks} = 2)$, or has filled surfaces $(y_{ijks} = 3)$ with $i = 1, \cdots, I$, $j = 1, \cdots, n_{i}$, $k = 1, \cdots, 28$, and $s = 1, \cdots, 4$ or $s = 1, \cdots, 5$, depending on the tooth type (incisor and canine teeth have four surfaces while molar and pre-molar teeth have five surfaces). We assume that the joint distribution of ${\bf y}_{i} = \big\{y_{ijks}\big\}$ follows a Potts model, given by
\begin{equation}\label{eq:auto}
\begin{split} %%%%%%%%%%%%%%%%%%%%%%%
f\Big({\bf y}_{i}&\mid\boldsymbol\theta_{p_i}, {\bf x}_{i}\Big)
= \frac{1}{\kappa\Big(\boldsymbol\theta_{p_i}\Big)}
\exp\bigg[\sum_{h=1}^5\psi_{p_i,h}s_h\big({\bf y}_i\big)\\
&+ \sum_{(j,k,s)}I\big(y_{ijks} = 2\big)\Big\{\alpha_{d_i} + \sum_{r=1}^6\beta_{r,d_i}z_{r,ij}\Big\}\\
&+\!\sum_{(j,k,s)}\!I\big(y_{ijks}\!=\! 3\big)\Big\{\alpha_{f_i}\! +\! \sum_{r=1}^6\beta_{r,f_i}z_{r,ij}\Big\} \\
&+\!\sum_{(j,k,s)}\Big\{I\big(y_{ijks} \!=\! 2\big)\!
^+\! I\big(y_{ijks}\! = \!3\big)\Big\}\!\sum_{q=1}^{k+2} \beta_{q,p_i} B_q(\pi_{ij})\bigg],\!\!
\end{split}
\end{equation}
where $\psi_{h,p_i}$ ($h = 1, \cdots, 5$) represents the five spatial interaction parameters; $\beta_{r,d_i}$ and $\beta_{r,f_i}$ measure the effects of covariates for the decayed and filled surfaces with $\alpha_{d_i}$ and $\alpha_{f_i}$ as the intercept, respectively, and $\beta_{q,p_i}$ is a regression coefficient of the quadratic B-spline for the inclusion probability. In this Potts model, $\kappa(\boldsymbol\theta_{p_i})$ is a doubly-intractable normalizing constant, where $\boldsymbol\theta_{p_i} = \big(\boldsymbol\psi_{p_i} = \{\psi_{h,p_i}\}, ~\alpha_{d_i}, ~\alpha_{f_i}, ~\boldsymbol\beta_{d_i} = \{\beta_{r,d_i}\}, ~\boldsymbol\beta_{f_i} = \{\beta_{r,f_i}\}, ~\boldsymbol\beta_{p_i}=\{\beta_{q,p_i}\}\big)$. With a one-to-one correspondence to $\psi_{h,p_i}$, the five spatial terms $s_h \big({\bf y}_i\big)$ are defined as follows:
\begin{itemize}
\item $s_1 \big({\bf y}_i\big) = \sum_{(j,k)}\sum_{s \neq 1} I\big(y_{ijks} = y_{ijk1} \big) I(x_{ijk} = 1)$, corresponding to $\psi_{1,p_i}$, which represents the associations between the occlusal surface and the other surfaces on the same tooth;
\item $s_2 \big({\bf y}_i\big) = \sum_{(j,k)}\sum_{s=4,5} \Big\{I\big(y_{ijks} = y_{ijk2}\big) + I\big(y_{ijks} = y_{ijk3}\big)\Big\} I\big(x_{ijk} = 1\big)$, corresponding to $\psi_{2,p_i}$, which represents the association between adjacent non-occlusal surfaces on the same tooth;
\item $s_3 \big({\bf y}_i\big) = \sum_{(j,k)}\sum_{m \sim k} I\big(y_{ijk2} = y_{ijm3}\Big) I\big(x_{ijk} = 1\big)I\big(x_{ijm} = 1\big)$, where $m \sim k$ represents $m = k - 1$ for $k = 2, \ldots, 7, 16, \ldots, 21$; and $m = k + 1$ with $k = 8, \ldots, 13, 22, \ldots, 27$. This corresponds to $\psi_{3,p_i}$, representing the association between the mesial and distal surfaces of adjacent teeth on the same jaw;
\item $s_4 \big({\bf y}_i\big) = \sum_{(j,k)}\sum_{s=1,4,5}\sum_{m \sim k} I\big(y_{ijks} = y_{ijms}\big) I\big(x_{ijkl} = 1\big)I\big(x_{ijm} = 1\big)$, where $m \sim k$ represent $m = k + 1$ for $k = 1, \ldots, 13, 15, \ldots, 27$ when $s = 2, 5$ and $m = k + 1$ for $k = 1, 2, 3, 11, 12, 13, 15, 16, 17, 25,$ $26, 27$ when $s = 1$. This corresponds to $\psi_{4,p_i}$, which represents the association between the adjacent occlusal, facial, and lingual surfaces of teeth on the same jaw; and
\item $s_5 \big({\bf y}_i\big) = \sum_{(j,k)}\sum_{o \leftrightarrow k} I\big(y_{ijk1} = y_{ijo1}\big) I\big(x_{ijk} = 1\big)I\big(x_{ijo} = 1\big)$, where $o \leftrightarrow k$ denotes the contacting teeth $o$ and $k$ on opposite jaws, corresponding to $\psi_{5,p_i}$, which represents the association between the occlusal surfaces of these teeth.
\end{itemize}

\subsection{Models for Non-ignorable Missingness}
\label{sec:non-ignorable}

As discussed in Section \ref{sec:Missing}, there are two types of nonresponses: NORD (i.e., the subject who did not provide poverty information) and NORP (i.e., the subject who missed the oral dental exam). Both are likely to induce non-ignorable missing data (i.e., NMAR). Let $r_{1,ij}$ and $r_{2,ij}$ be nonresponse indicators signifying whether individual $j$ in PSU $i$ is NORD or NORP, respectively, $i = 1, \cdots, I$ and $j = 1, \cdots, n_{i}$. If a subject is NORP (i.e., $r_{1,ij}=0$), the subject's poverty status is missing; and if a subject is NORD (i.e., $r_{2,ij}=0$), the subject's oral health data (including tooth and surface outcomes and two covariates sealant and fluorosis) are missing. Let $v_{k, ij}$ generically denote the variables that are missing due to $r_{k,ij}=0, k=1, 2$, and $\bf{u}_{ij}$ generically denotes other completely observed covariates.
We model the non-ignorable missing data using the selection model \citep{Little:2002, Little:2008p409} as follows:
\begin{equation}\label{eqn:selection} %%%%%%%%%%%%%%%%%%%%%%%
\begin{split}
f\big(r_{k,ij}, v_{k,ij} &\mid {\bf u}_{k, ij}, \boldsymbol\gamma_{k,i}, \boldsymbol\vartheta_{k,i}\big) = \\
& f\big(r_{k,ij} \mid v_{k,ij}, {\bf u}_{k, ij}, \boldsymbol\vartheta_{k,i}\big)
\times f\big(v_{k,ij} \mid {\bf u}_{k,ij}, \boldsymbol\gamma_{k,i}\big), \quad k = 1, 2,    
\end{split}
\end{equation}
where $\boldsymbol\vartheta_k$ and $\boldsymbol\gamma_k$ are the model parameters. 

\subsubsection{Selection Model for NORP} 
\label{sec:MissingPoverty} 
For NORP, we assume that $f\big(r_{1,ij} \mid{\bf u}_{1, ij}, v_{1,ij}, \boldsymbol\vartheta_{1,i}\big)$ in Eqn. \ref{eqn:selection} for the missingness of poverty follows: 
\begin{equation}\label{eq:poverty.missing} %%%%%%%%%%%%%%%%%%%%%%%%
f\big(r_{1,ij} \mid {\bf u}_{1,ij}, v_{1,ij}, \boldsymbol\vartheta_{1,i} \big) 
\sim \text{Bernoulli} \bigg\{\mbox{logit}\Big(\boldsymbol\vartheta_{1,i}^{T} {\bf u}_{1,ij} + \vartheta_{r_1,i} v_{1,ij}\Big)\bigg\},
\end{equation}
where ${\bf u}_{1,ij}=\big(1, {\bf z}_{ij}, s({\bf x}_{ij}), {\bf s}_h({\bf y}_{ij})\big)^T$; ${\bf z}_{ij}$ is the baseline individual-level covariates other than poverty; $s({\bf x}_{ij})$ contains the statistics representing the spatial interaction at the tooth level; and ${\bf s}_h({\bf y}_{ij})$ contains statistics for the five spatial interactions at the surface level for individual $j$ in the PSU $i$; and $\boldsymbol\vartheta_{1,i}$ contains the corresponding regression coefficients. $f\big(v_{1,ij} \mid {\bf u}_{1,ij}, \boldsymbol\gamma\big)$ in Eqn. \ref{eqn:selection} for modelling the binary poverty is a logistic regression model:
\begin{equation*} \label{eq:poverty}
\mbox{logit}\Big\{\Pr(v_{1,ij}=1) \mid {\bf u}_{1,ij}, \boldsymbol\gamma_{1,i}\Big\} = \boldsymbol\gamma_{1,i}^T {\bf u}_{1,ij}.
\end{equation*}

\subsubsection{Selection model for NORD} 
\label{sec:MissingOutcome}

To modeling the non-ignorable missingness of NORD, we assume that
$f\big(r_{2,ij} \mid {\bf u}_{2,ij}, r_{1,ij}, \boldsymbol\vartheta_{2,i}\big)$ in Eqn. \ref{eqn:selection} follows
\begin{equation*}\label{eq:tooth.missing}
f\Big(r_{2,ij} \mid {\bf u}_{2,ij}, r_{1,ij}, \boldsymbol\vartheta_{2,i}\Big)
\sim \mbox{Bernoulli} \bigg\{\mbox{logit} \Big(\boldsymbol\vartheta_{2,i}^{T} {\bf u}_{2,ij} +
\vartheta_{r_2,i} r_{1,ij} \Big)\bigg\}
\end{equation*}
where ${\bf u}_{2, ij} = \big(1, {\bf z}_{ij}, s({\bf x}_{ij}), {\bf s}_h({\bf y}_{ij})\big)^T$; ${\bf z}_{ij}$ includes all the baseline individual-level covariates, including sealant and fluorosis; and $s({\bf x}_{ij})$ and ${\bf s}_h({\bf y}_{ij})$ are defined in the same way as for Eqn. \ref{eq:poverty.missing}. Note that we also included $r_{1,ij}$,  the poverty missing indicator, in the regression model, as $r_1$ and $r_2$ might be correlated. 

For NORD, both outcome variables (i.e., tooth outcome $x_{ijks}$ and surface outcome $y_{ijks}$) and two covariates (i.e., sealant and fluorosis) are missing. For dental outcomes $x_{ijks}$ and $y_{ijks}$, $f\big(v_{2,ij} \mid {\bf u}_{2,ij}, \boldsymbol\gamma_{2,i}\big)$ in Eqn. \ref{eqn:selection} are provided by the Potts models (i.e., Eqn. \ref{eq:potts} and \ref{eq:auto}). For sealant (a binary indicator variable taking values of yes or no), denoted as $v_{s,ij}$, we assume $f\big(v_{s,ij} \mid {\bf u}_{s,ij}, \boldsymbol\gamma_{s,i}\big)$ in Eqn. \ref{eqn:selection} follows a logistic regression model, 
\[ \mbox{logit}\Big\{Pr\big(v_{s,ij} = 1\big) \mid {\bf u}_{s,ij}, \boldsymbol\gamma_{s,i}\Big\} = \boldsymbol\gamma_{s,i}^T {\bf u}_{s,ij}, \] 
where ${\bf u}_{s, ij}=\big(1, {\bf z}_{ij}, s({\bf x}_{ij}), {\bf s}_h({\bf y}_{ij})\big)^T$ and ${\bf z}_{ij}$ represents the baseline individual-level covariates excluding sealant. 
Fluorosis used in our model is defined as the average fluorosis value for all the present teeth. Given that an individual usually has about 28 teeth, it is reasonable to assume that fluorosis, denoted as $v_{f,ij}$, is approximately normal and follows a linear regression model,
\[ v_{f,ij} \mid {\bf u}_{f,ij}, \boldsymbol\gamma_{f,i} = N(\boldsymbol\gamma_{f,i}^T{\bf u}_{f,ij}, \boldsymbol\varphi_i^2), \]
where ${\bf u}_{f, ij}=\big(1, {\bf z}_{ij}, s({\bf x}_{ij}), {\bf s}_h({\bf y}_{ij})\big)^T$, ${\bf z}_{ij}$ represents the baseline individual-level covariates excluding fluorosis and sealant, and $\boldsymbol\varphi_i^2$ is the variance parameter. 
%{\color{red}check if $\tau^2$ has been used somewhere}

\subsection{Hierarchical Modelling for Small Area Estimation}\label{sec:SmallArea}
As described in Section \ref{sec:smallarea}, the sample size in some PSUs is small and cannot provide sufficient information to reliably estimate some model parameters in each PSU separately. We employ a hierarchical modelling framework as a small area estimation technique to borrow information across PSUs. In addition, the hierarchical structure also naturally accounts for the multistage sampling scheme \citep{skinner1989analysis}. The exploratory results in Figure 3 in Section B of Supplement Materials suggest that it is plausible to use normal distributions as the prior for the model parameters in the outcome measurement models. Specifically, for each of the regression coefficient $k$ from the measurement model in PSU $i$, we define
$\theta_{k,i} \sim \mbox{N}\big(\delta_{\theta_k}, \sigma_{\theta_k}^2\big)$, where
$\delta_{\theta_k} \sim \mbox{N}\big(\lambda_{\theta_k}, \tau^2_{\theta_k}\big);\;
\sigma^2_{\theta_k} \sim \mbox{IG}(a_{\theta_k},b_{\theta_k})$.

We also applied similar priors and hyper-priors to the parameters from the selection models and the imputation models for attributing missing values in the covariates; 
$\gamma_{k,i} \sim \mbox{N}\big(\delta_{\gamma_k}, \sigma^2_{\gamma_k}\big)$ 
where 
$\delta_{\gamma_k} \sim \mbox{N}\big(\lambda_{\gamma_k}, \tau^2_{\gamma_k}\big);\;
\sigma^2_{\gamma_k} \sim \mbox{IG}(a_{\gamma_k},b_{\gamma_k});\;$
and 
$\vartheta_{k,i}\sim \mbox{N}\big(\delta_{\vartheta_k}, \sigma_{\vartheta_k}^2\big)$ 
where $\delta_{\vartheta_k} \sim \mbox{N}\big(\lambda_{\vartheta_k}, \tau^2_{\vartheta_k}\big);\;
\sigma^2_{\vartheta_k} \sim \mbox{IG}\big(a_{\vartheta_k},b_{\vartheta_k}\big).$
We set $\lambda_{*_k}$ ($*\in\{\theta,\delta,\vartheta\}$) at 0.5 for the spatial interaction parameters and at 0 for the regression coefficients associated with other covariates, $\tau_{*_k} = 5$, and $a_{*_k}= b_{*_k} = 0.001$ for all $k$. 

In principle, we could introduce more model hierarchies to mirror each of the sampling stages (i.e., sampling segments within PSU, sampling households within the segment, and sampling subjects within the household). We did not do this, because the published NHANES data do not contain the segment and household identifiers (i.e., there is no information to identify which segment and household a specific subject belongs to). As the sampling weights somewhat already contain the segment and household sampling information (see Section \ref{sec:data_describe}), ignoring these sampling procedures might have little impact on the inference.

\section{Results}
We apply the method described in Section \ref{sec:Methdology} to the NHANES dental data. We employed the noisy Monte Carlo sampler to generate posterior samples. Our MCMC run consisted of 30,000 iterations, with 20 auxiliary samples for each iteration to evaluate the normalizing constants ratios. We discarded the first 5,000 iterations for the burn-in process, and used a thinning of 5 iterations to collect 5,000 samples from the remaining iterations. The main results are summarized below.

\subsection{Covariate effects on dental caries}

Table \ref{tab:covariate} summarizes the posterior means and 95\% highest posterior density (HPD) intervals for the parameters from Potts models for the tooth and surface outcomes, quantifying the effects of various covariates on the carious conditions (i.e., missing teeth due to the disease, missing teeth due to the other reason, decayed and filled surfaces). 

\begin{table}[htbp]
\centering
\begin{tabular}{llcrr} \hline
Covariate & Outcome Condition & Posterior mean & \multicolumn{2}{c}{95\% HPD} \\ \hline
Intercept &
   Missing (Disease) & -1.5313 & (-2.1193, & -0.8921) \\ 
 & Missing (Other)   & -1.9772 & (-2.7531, & -1.1871) \\ 
 & Decayed &  1.3158 & (-0.0331, &  2.5716) \\ 
 & Filled  &  3.6222 & ( 2.4099, &  4.7270) \\ \hline
Gender & 
   Missing (Disease) & -0.0622 & (-0.4962, &  0.3773) \\ 
 & Missing (Other)   & -0.3947 & (-0.9575, &  0.1362) \\ 
 & Decayed &  0.3134 & ( 0.0305, &  0.6003) \\
 & Filled  & -0.1121 & (-0.2043, & -0.0153) \\ \hline
Poverty &
   Missing (Disease) & -0.1366 & (-0.5911, &  0.2938) \\ 
 & Missing (Other)   &  0.9306 & ( 0.2354, &  1.5709) \\ 
 & Decayed & -0.2239 & (-0.5713, &  0.1870) \\ 
 & Filled  &  0.4793 & ( 0.2291, &  0.7615) \\ \hline
Race (White) &
   Missing (Disease) & -1.2013 & (-2.2308, & -0.2440) \\ 
 & Missing (Other)   &  1.1509 & ( 0.2075, &  2.1915) \\ 
 & Decayed & -0.5098 & (-1.3837, &  0.3716) \\
 & Filled  &  0.1112 & (-0.5248, &  0.7493) \\ \hline
Race (Black) &
   Missing (Disease) &  0.3192 & (-0.7758, &  1.3666) \\ 
 & Missing (Other)   & -0.7758 & (-1.9332, &  0.4796) \\ 
 & Decayed &  0.0158 & (-0.9314, &  0.9272) \\
 & Filled  &  0.0944 & (-0.7201, &  0.9193) \\ \hline
Sealant &
   Missing (Disease) & -2.9429 & (-3.8822, & -1.9144) \\ 
 & Missing (Other)   & -1.2722 & (-2.2364, & -0.3144) \\ 
 & Decayed & -2.9894 & (-3.9718, & -2.1118) \\ 
 & Filled  & -1.1967 & (-1.7942, & -0.6432) \\ \hline
Fluorosis &
   Missing (Disease) & -2.2475 & (-2.7493, & -1.7687) \\ 
 & Missing (Other)   & -1.9646 & (-2.4873, & -1.4677) \\ 
 & Decayed & -0.7888 & (-1.0625, & -0.4959) \\ 
 & Filled  & -0.1653 & (-0.2457, & -0.0923) \\  \hline
\end{tabular}
\caption{
\label{tab:covariate}
Posterior means and 95\% HPD intervals of the pooled covariate-effect parameters.}
\end{table}

For example, the parameter corresponding to gender represents the difference between female and male in the log odds of having a missing tooth, either due to dental disease (log odds = $-0.0622$) or other reasons (log odds = $-0.3947$) vs. no missing teeth in the Potts model for the tooth outcome, conditional on the other covariates and spatial referencing for that spatial location remaining the same. Similarly, in the Potts model for the surface outcome, the parameter corresponding to gender represents the difference between female and male in the log odds of having a decayed (log odds = $0.3134$) or filled surface (log odds = $-0.1121$) vs. a healthy surface at the same spatial location, while other covarites and spatial associations remaining same.
%{\color{red} break this "For example...." sentence, too long, and point out which number in table 4 corresponds to each of the statements}

The effects of other covariates can be interpreted in a similar fashion. If the 95\% HPD interval of a parameter does not include 0, we could claim the covariate corresponding to that parameter has a substantial effect on the caries outcomes.

The posterior means according to sealant and fluorosis were all negative, suggesting that having sealants and fluorosis reduces the risks of having dental caries overall, as expected. The result also shows that females are less likely to have filled surfaces (log odds $= -0.1121$) and more likely to have decayed surfaces (log odds $= 0.3134$). People above the poverty line have increased odds of losing teeth from other reasons (log odds $= 0.9272$) and filling surfaces after decayed (log odds $= 0.4793$). The non-Hispanic white population tends to have more missing teeth from other reasons (log odds $= 1.1509$) and has less missing teeth due to disease (log odds $=-1.2013$). This is compared to the reference race group (most of which are Hispanic), while the differences in caries outcomes between non-Hispanic black and Hispanic populations are insignificant.

The two intercept terms from the two Potts model can be interpreted as the conditional log odd-ratios of having missing teeth due to disease, and those due to other reasons with non-missing teeth as the reference, as well as the conditional log odd-ratios of having decayed or filled surfaces with healthy surfaces as the reference, respectively. The results suggest that having missing teeth, due to the dental disease (log odds = $-1.5313$) and due to other reasons (log odds = $-1.9772$), are less likely than preserving teeth among survey participants. Missing teeth from dental disease is more common than due to other reasons, and filled (log odds = $3.6222$) and decayed (log odds = $1.3158$) surfaces are more common than healthy surfaces. 

\subsection{Spatial Association Parameters}

Table \ref{tab:spatial} summarizes the posterior means and 95\% HPD intervals of the spatial association parameters. Usually, in the Potts model specification, a value of 1.0 for the spatial association parameters $\psi$ amounts to a very high degree of associations \citep{Green:2002p1055}. 

\begin{table}[htbp]
\centering
\begin{tabular}{lrrr} \hline
  parameter & post. mean & \multicolumn{2}{c}{95\% HPD} \\ \hline
Tooth      & 0.6074 & ( 0.5663, &  0.6528) \\
Type-A$_1$ & 0.0964 & ( 0.0821, &  0.1111) \\ 
Type-A$_2$ & 1.2626 & ( 1.2262, &  1.3007) \\ 
Type-B$_1$ & 0.8711 & ( 0.8164, &  0.9284) \\ 
Type-B$_2$ & 0.6440 & ( 0.6134, &  0.6746) \\ 
Type-C &  0.0003 & ( 0.0000, &  0.0010) \\ \hline
% B-spline   & -2.8440 & (-3.8213, & -1.7804) & B-spline   & -4.7760 & 
%(-6.0326, & -3.2962) \\  
% parameters & -3.7110 & (-4.5675, & -2.8173) & parameters & -4.7615 & 
%(-5.9385, & -3.4051) \\  
% (tooth) & -4.0620 & (-4.8285, & -3.3917)    & (surface)  & -5.1071 & 
%(-6.2259, & -3.8477) \\  
%         & -3.1778 & (-4.3213, & -2.0304)    &            & -4.6338 & 
%(-5.8858, & -3.2962) \\ 
%         & -1.5818 & (-3.1184, & -0.0966)    &            & -2.1818 & 
%(-3.5910, & -0.8189) \\ \hline
\end{tabular}
\caption{
\label{tab:spatial}
Posterior mean estimates and 95\% HPD intervals of the pooled spatial association parameters and pooled B-spline parameters.}
\end{table}

The estimate at the tooth level was 0.6074, implying a moderate-high level of association. At the surface level, the posterior estimates of five spatial association parameters suggest the strongest association was between adjacent non-occlusal surfaces on the same tooth (Type-A$_2$), followed by the association between the mesial and distal (contacting) surfaces of adjacent teeth on the same jaw (Type-B$_1$), and the association between the adjacent occlusal, facial, and lingual (non-contacting) surfaces of teeth on the same jaw (Type-B$_2$), while that of contacting occlusal surfaces on opposite jaws (Type-C) and that between the occlusal surface and the other surfaces on the same tooth (Type-A$_1$) are negligible. To summarize, there exist high associations between non-occlusal surfaces, while those with occlusal surfaces are less likely. In other words, the caries outcomes of the occlusal surfaces are unlikely to influence those of non-occlusal surfaces. This observation is consistent with the results in \citet{Jin:2016p884}. 

\begin{table}[htbp]
\centering
\scalebox{0.85}{
\begin{tabular}{lrrrlrrr}
\hline\hline
 \multicolumn{7}{c}{Model for sealant} \\
  \hline
Covariate & Estimates & \multicolumn{2}{c}{95\% HPD} &
Associations & Estimates & \multicolumn{2}{c}{95\% HPD} \\ 
  \hline
Intercept    & -3.2172 & (-3.5479, & -2.8764) & Tooth      &  0.1155 & (-0.2376, &  0.4417) \\ 
Gender       & -0.4317 & (-0.7485, & -0.1077) & Type-A$_1$ &  0.4494 & ( 0.1284, &  0.7746) \\ 
Poverty      &  0.4514 & (-0.0531, &  1.0134) & Type-A$_2$ &  0.9076 & ( 0.3809, &  1.4436) \\ 
Race (White) &  1.1951 & ( 0.8052, &  1.5926) & Type-B$_1$ & -0.0779 & (-0.6041, &  0.4758) \\ 
Race (Black) & -0.6521 & (-1.3502, & -0.1183) & Type-B$_2$ & -0.0620 & (-0.5935, &  0.4632) \\
Fluorosis    & -0.1602 & (-0.4615, &  0.1243) & Type-C     & -0.4485 & (-0.7107, & -0.1800) \\  \hline
 \multicolumn{7}{c}{Model for fluorosis} \\
  \hline
Covariate & Estimates & \multicolumn{2}{c}{95\% HPD} &
Associations & Estimates & \multicolumn{2}{c}{95\% HPD} \\ 
  \hline
Intercept    & -0.1856 & (-0.2335, & -0.1383) & Tooth      &  0.0013 & (-0.0372, &  0.0391) \\ 
Gender       &  0.0081 & (-0.0493, &  0.0603) & Type-A$_1$ & -0.0545 & (-0.0991, & -0.0127) \\ 
Poverty      &  0.0182 & (-0.0500, &  0.0893) & Type-A$_2$ &  0.0985 & ( 0.0324, &  0.1606) \\ 
Race (White) & -0.0054 & (-0.0667, &  0.0578) & Type-B$_1$ &  0.0302 & (-0.0423, &  0.1031) \\ 
Race (Black) &  0.1270 & ( 0.0375, &  0.2072) & Type-B$_2$ &  0.2091 & ( 0.1398, &  0.2817) \\
Fluorosis    &  0.0270 & (-0.0567, &  0.1168) & Type-C     &  0.0300 & (-0.0134, &  0.0697) \\  \hline
\multicolumn{7}{c}{Model for poverty} \\
\hline
Covariate & Estimates & \multicolumn{2}{c}{95\% HPD} &
Associations & Estimates & \multicolumn{2}{c}{95\% HPD} \\ 
  \hline
Intercept    &  0.9320 & ( 0.7996, &  1.0640) & Tooth      & -0.0636 & (-0.1929, &  0.0643) \\ 
Gender       &  0.4688 & ( 0.2770, &  0.6245) & Type-A$_1$ & -0.7260 & (-0.8333, & -0.6310) \\ 
Race (White) &  0.9748 & ( 0.7452, &  1.2194) & Type-A$_2$ &  0.5259 & ( 0.3299, &  0.7704) \\ 
Race (Black) &  0.0793 & (-0.2537, &  0.4213) & Type-B$_1$ & -0.1564 & (-0.4330, &  0.1571) \\ 
Sealant      & -0.1364 & (-0.4604, &  0.1247) & Type-B$_2$ &  0.0380 & (-0.3203, &  0.2967) \\
Fluorosis    & -0.0418 & (-0.1466, &  0.0594) & Type-C     &  0.2801 & ( 0.1372, &  0.4384) \\   \hline
\end{tabular}
}
\caption{
\label{tab:impu}
Posterior means and 95\% HPD intervals of the pooled parameters in the imputation model for sealant, fluorosis and poverty}
\end{table}

\subsection{Parameters in Imputation Models for Sealant, Fluorosis and Poverty}

%Table \ref{tab:sealant}, Table \ref{tab:fluorosis}, and Table \ref{tab:poverty} 

Table \ref{tab:impu} summarizes the posterior means and 95\% HPD intervals of the parameters from the models for sealant, fluorosis, and poverty, given other covariates, respectively. These parameters measure the effects of individual-level covariates and spatial association among teeth and surfaces on the tendency of having sealant in the molar teeth, the fluorosis level in teeth, and the likelihood of above poverty line, respectively. The other parameters in the selection models from modelling the non-ignorable missingness on these covariates are summarized in %Section C of the 
Section C of Supplement Materials.
\begin{figure}[!htb]
\begin{center}
\begin{tabular}{cc}
(a) Tooth Absence Due to Disease & 
(b) Tooth Absence from Other Reasons \\
\includegraphics[trim={0.1cm 0.5cm 0.1cm 1.7cm},clip,width=0.45\textwidth] {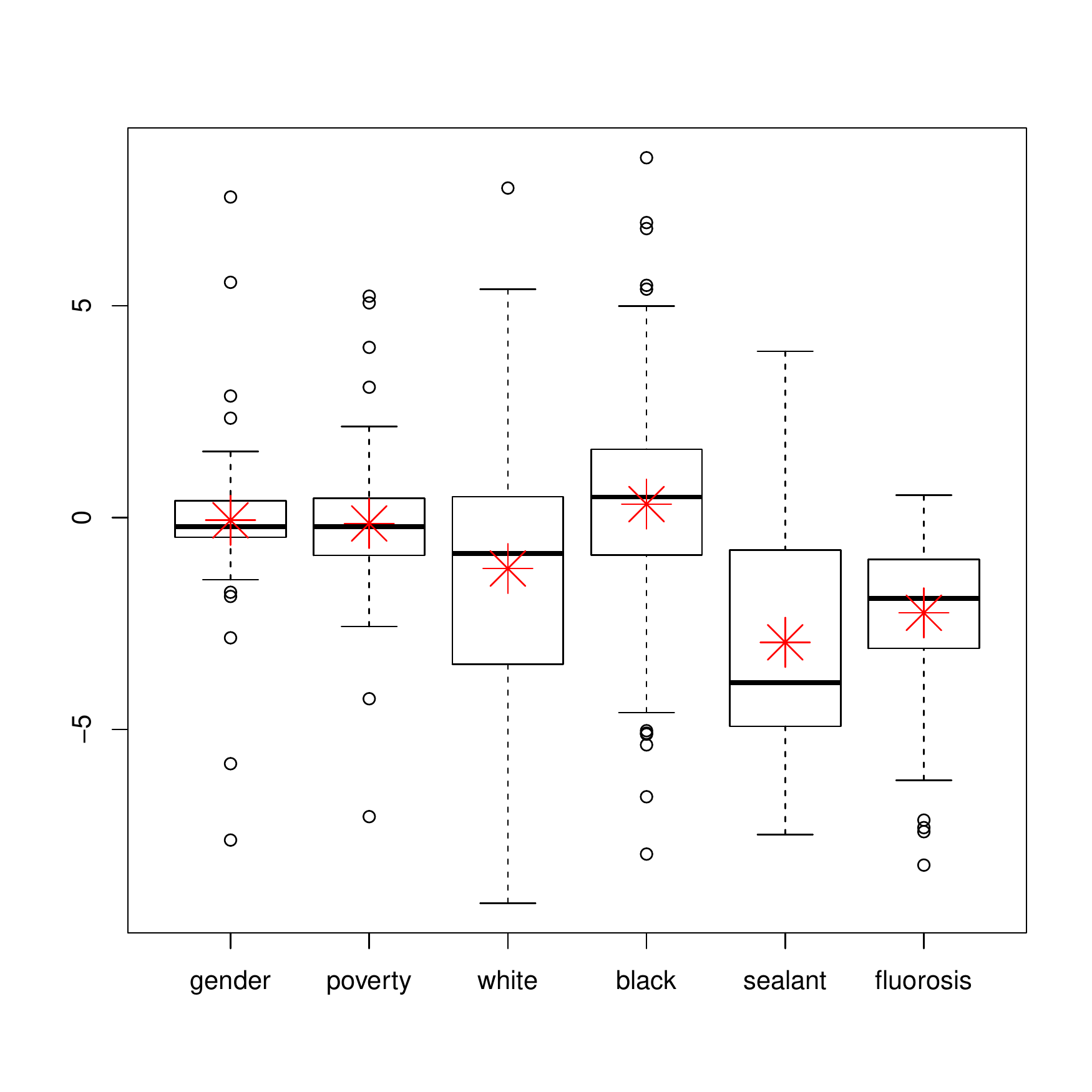} & 
\includegraphics[trim={0.1cm 0.5cm 0.1cm 1.7cm},clip,width=0.45\textwidth] {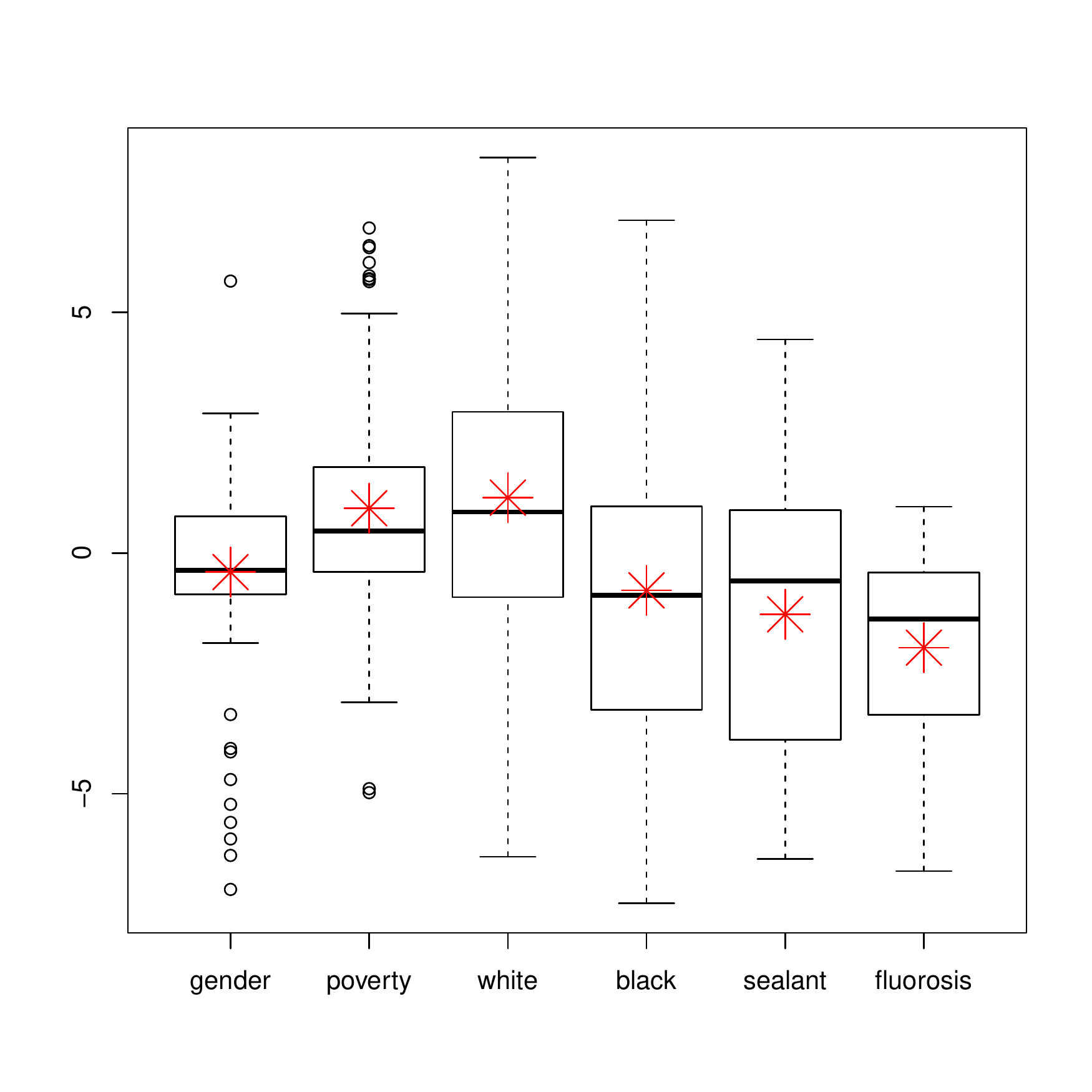} \\
(c) Decayed Surfaces & 
(d) Filled Surfaces \\
\includegraphics[trim={0.1cm 1cm 0.1cm 1.7cm},clip,width=0.45\textwidth] {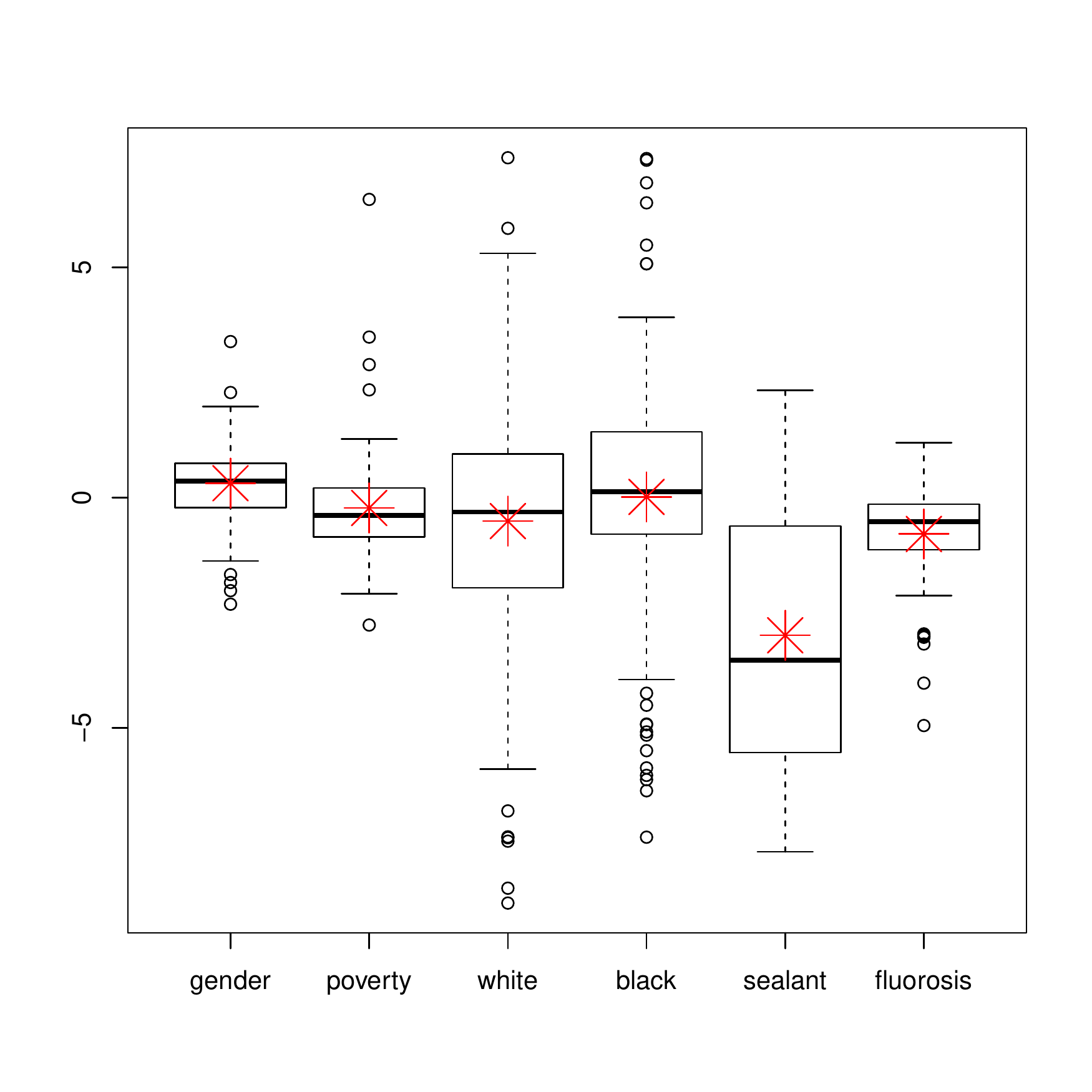} &
\includegraphics[trim={0.1cm 1cm 0.1cm 1.7cm},clip,width=0.45\textwidth] {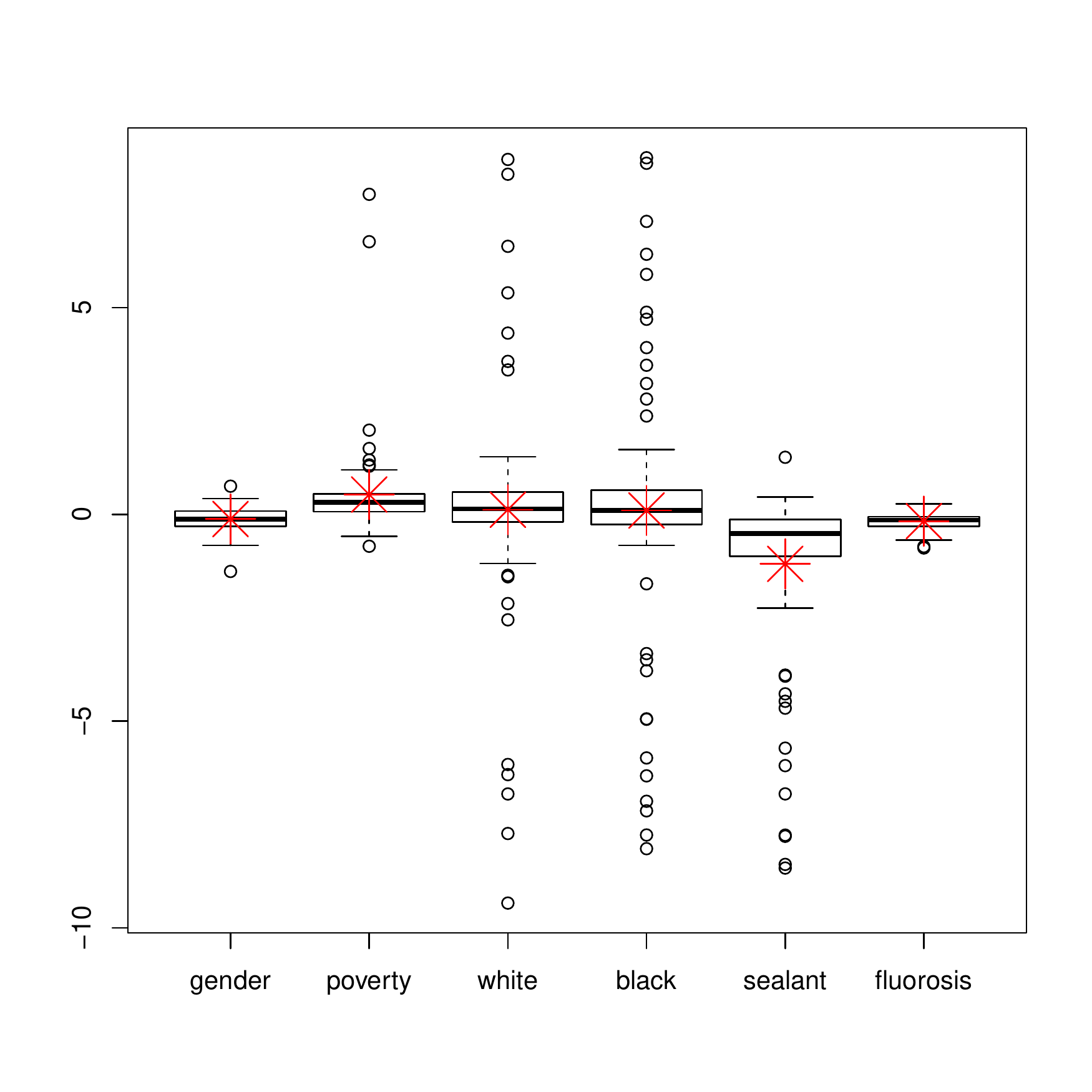} \\
\end{tabular}
\caption{
Boxplots for PSU-level posterior mean estimates of covariate-effect parameters at the tooth and surface levels.
The red marks in boxplots represent the pooled posterior mean estimates of covariate-effect parameters.
}
\label{fig:small_area}
\end{center}
\end{figure}

The results in Table \ref{tab:impu} suggest that the non-Hispanic white population tends to have preventive sealant treatments for their molar teeth, compared to the reference race group (log odds $= 1.1951$); and females are less likely to have sealants compared to males (log odds $= -0.4317$). The type-A spatial associations (spatial associations within a single tooth) show a positive relationship, while the type-C spatial association (spatial association of contacting occlusal surfaces on opposite jaws) has a negative relationship with sealants. 
%The results in Table \ref{tab:impu} 
In terms of fluorosis, the results suggest that the non-Hispanic black population tends to have a higher level of fluorosis in their teeth, compared to the reference race group (log odds$= 0.1270$). While the existence of Type-A$_2$ and Type-B$_2$ spatial association tends to promote the fluorosis level, while the Type-A$_1$ association tends to decrease the fluorosis-level in teeth. The results in Table \ref{tab:impu} also suggest that females as compared to males (log odds $=0.4688$) and non-Hispanic whites as compared to the Reference Race group (log odds $=0.9748$) are more likely to be above the poverty line.

\subsection{Small Area Estimation Results} 
The Bayesian hierarchical spatial model we proposed for small area estimation helps to stabilize the PSU-level parameter estimates, which could be otherwise unreliable or even impossible due to the small sample sizes in some PSUs. To check the PSU-level parameters are well estimated under our model, we draw the box plots for PSU-level posterior mean estimates of the covariate-effect parameters, as shown in Figure \ref{fig:small_area}. Red marks in the box plots represent the overall posterior mean estimates of these parameters. Figure \ref{fig:small_area} shows that all PSU-level parameters are reliably estimated under our modeling framework. Numerical results at the tooth and surface levels are also summarized in 
Section D of the Supplement Materials. %Section D of the supplementary material 

\section{Discussion}
In this paper, we proposed a new Bayesian hierarchical spatial model for small-area estimation with non-ignorable nonresponse, and applied it analyze dental caries outcomes collected in NHANES from participants aged from 20 to 34. 
%To analyze the dental outcomes in each PSU, we refined a Bayesian spatial hierarchical model proposed by \citet{Jin:2016p884}, which closely resembles the caries evolution process in humans. At the tooth level, a Potts model was used to model the trinary probability of a tooth being present, missing due to dental disease, or missing due to other reasons. At the surface level, conditional on the non-missing tooth, we model the probability of a decayed, filled, or healthy surface via a second Potts model. To take into account the effect of the complex survey design on the dental outcomes, we employed the B-splines on the sampling weights in the outcome models. To handle the sparse information on some model parameters for some PSUs, we used the Bayesian hierarchical framework to borrow information across the PSUs in the survey. We exploited the selection model to model the non-ignorable missingness, both in the covariates and in the outcome variables. We combined the data augmentation method and the noisy Monte Carlo sampler to estimate parameters from the proposed mode. 
The analysis results suggest that there exists strong spatial associations between teeth, between adjacent non-occlusal surfaces on the same tooth, and between the contacting and non-contacting surfaces of adjacent teeth on the same jaw. The dental hygienic factors of fluorosis and sealants reduce the risk of having dental diseases. 
%Females are more likely to have decayed surfaces, but are less likely have filled surfaces. If respondents are above the poverty line, then they tend to fix their cavities. Non-Hispanic white populations lose less teeth due to disease, compared to other race groups (mostly Hispanic), whereas there are insignificant differences in caries outcomes between non-Hispanic black populations and other races. The imputation models suggests that non-Hispanic whites get more preventive sealant treatment for their molar teeth, whereas females tend to have less sealant treatment. In addition, non-Hispanic blacks tend to have fluorosis in their teeth; and females and non-Hispanic whites are more likely to be above the poverty line. 

As an alternative to the proposed measurement model, one may also use a multinomial framework in SGLM that uses a latent Gaussian Markov random field to model spatial dependence. We chose to use the Potts model, because spatial dependence can be easily interpreted in this tool, while choosing the cut-off values for the latent Gaussian Markov random fields is often challenging in the SGLM because dental outcomes are nominal variables.

%If there are many covariates collected from the participants in a survey, variable selection can be incorporated into our measurement model, which would need to take into account the doubly-intractable normalizing constants \citep{Caimo:2013, Bouranis:2018}.

%Though we developed the measurement model in the framework of a dental survey data set, it can be easily extended to other survey data, which include general bivariate spatial outcomes with mixed binary and multinomial outcomes in their measurements.  

\bibliography{Sampling}
\end{document}